# Fractal kinetics versus fractional derivative kinetics

Francois Brouers[a], Tariq J. Al-Musawi[b]

[a] Faculty of Applied Sciences, Liège University, Belgium, E-mail address: fbrouers@ulg.ac.be
[b] Faculty of Engineering, Isra University, Amman, Jordan, E-mail address: tariqjwad@yahoo.com

**Abstract**

This study presents a detailed comparison of the two most popular fractal theories used in the field of kinetics sorption of pollutants in porous materials: the Brouers-Sotolongo model family of kinetics based on the BurrXII statistical distribution and the fractional kinetics based on the Riemann-Liouville fractional derivative theory. Using the experimental kinetics data of several studies published recently, it can be concluded that, although these two models both yield very similar results, the Brouers-Sotolongo model is easier to use due to its simpler formal expression and because it enjoys all the properties of a well-known family of distribution functions. We use the opportunity of this study to comment on the information, in particular, the sorption strength, the half-life time, and the time dependent rate, which can be drawn from a complete analysis of measured kinetics using a fractal model. This is of importance to characterize and classify sorbent-sorbate couples for practical applications. Finally, a generalization form of the Brouers-Sotolongo equation is presented by introducing a time dependent fractal exponent. This improvement, which has a physical meaning, is necessary in some cases to obtain a good fit of the experimental data.

## 1. Introduction

One can find nowadays a large number of works, both experimental and theoretical, dealing with the problem of vital importance for environment and public health. These efforts which are intensively discussed in the literature are essentially concerned with the problems of the removal of hazardous materials from water and wastewater to acceptable limits using a number of treatment techniques. Among them, the sorption techniques (adsorption; biosorption, and ion exchange) have been found to be efficient to eliminate harmful or potentially deadly pollutants. The degree of pollutants sorption is strongly depended on the surface characteristics of the chosen sorbent such as the geometry and the sorption energy distribution, as well as, the physical and chemical nature of the interaction between the pollutant molecules and the sorbing sites. Sorption kinetics describes time-dependent solute uptake which, in turn, controls the residence time of sorbate uptake at the solid–solution interface. This process is a highly complex phenomenon and consists of three steps which depend on the nature of the sorbent-sorbate couple and the external conditions (temperature, pH, concentration of pollutants, etc.). Many natural and artificial sorbents were tested for their applicability in the wastewater systems but only a little number was found to be practically usable. In this field, an intellectual gap has appeared in the course of years between the many people involved in the practical issues of creating new performing sorbents using the rapid progress in materials sciences and the theorists taking advantage of new mathematical software and filling scientific journals with statistical functions more and more sophisticated and practically unsuitable for practitioners in the field of sorption (**Handique and Chakraborty, 2017; Mdlongwa et al., 2017; Yari and Tondpour, 2017**). Part of this gap is due to the fact that many experimentalists still use routine and old-fashioned methods to interpret their results as well as many theorists forget that the macroscopic world cannot be described by simply extrapolating and that the number of pertinent variables should be limited and have some clear physical interpretation. This is an old philosophical problem going back to Occam's razor principle: *entia non sunt multiplicanda praeter necessitatem.*

Therefore, the present work is written with the objective to reduce this gap and to present a method which takes advantage of new development in statistical and stochastic theories. We will do that by reviewing, commenting, and comparing some recent published kinetics works.

The philosophy of our approach is to define and calculate the macroscopic quantities characterizing the sorption with mathematically well-defined functions containing a minimum of parameters. To interpret macroscopic data, we have to approach the problem with a mixture of mathematical rigor and empirical flair and to leave the belief that the use of microscopic or mesoscopic precise models, with the help of supercomputers, might be able to describe macroscopically complex systems. The reason is simply that the experience has shown that several different microscopic models can yield the same macroscopic equations. The best example is the Weibull function which can represent a number of different and unrelated physical systems with a variety of static and dynamic interactions.

For practical applications it is important to know to the rate of sorption and the maximum quantity of pollutant removed from aqueous solutions. The measure of the kinetics results from successions of averages over several scale levels can only give macroscopic quantities which are the ones used for practical applications. In the field of sorption kinetic, two methods based on the concept of fractality have been proposed to achieve this goal. First, the Brouers-Sotolongo (*BSf(n,a)*) (also called *BSW(t)* in several previous references) fractal kinetics introduced in references (**Brouers and Sotolongo-Costa,**

**2006; Brouers, 2014; Al-Musawi et al., 2017**) is based on the solutions of the Burr XII statistical function (**Burr, 1942; Singh and Maddala, 1976; Rodriguez, 1977**) and has been used in a number of papers with success in the field of complex sorbing systems mostly in aqueous phase (**Hamissa et al., 2007; Kesraoui et al., 2016; Ncibi et al., 2009; Figaro et al., 2009; Hamissa et al., 2013**). This model interpolates between the first- and second-order kinetics models. However, more importantly, it introduces not only a fractional order ($n$) but also a fractal time parameter ($a$) that characterizes the rate constant variations in time. On the hand, more recently the fractional derivative theory has been applied in the field of relaxation and sorption (**Tomczak et al., 2013; Kaminski et al., 2016**). They both rely on the concept of time fractality and include memory non-Markovian effects. We will compare these two methods theoretically and experimentally. Finally, we will take the opportunity of this work to introduce a further improvement of the *BSf(n, α)* model, by assuming a time variation of the fractal exponent to account the change of the physical conditions during the evolution of the sorbing process.

## 2. Materials and methods

### 2.1 Experimental data

In the present study, seven sets of experimental kinetic data of: tetracycline adsorption onto magnesium oxide-coated lightweight expanded clay aggregate (*set 1*), indigo carmine adsorption onto activated carbon without and with alternating current (*set 2 and 3,* respectively); cyanide onto LTA-zeolite (*set 4*); copper recovery from the Chilean mineral deposit by flotation (*set 5*); water vapor adsorption in cement (*set 6*), and fluorine adsorption onto clay (*set 7*) were analyzed. The experimental procedure details of these kinetics data are found in (**Al-Musawi et al., 2017; Kesraoui et al., 2016; Noroozi et al., 2017; Vinnett et al., 2015; Saeidpour and Wadsö, 2015; Brouers and Guiza, 2017**).

## 2.2 Mathematical models

### 2.2.1 The Brouers-Sotolongo fractal kinetic model

The original Burr XII distribution (**Burr, 1942; Singh and Maddala, 1976**; **Brouers and Sotolongo-Costa, 2006; Brouers, 2014a; Al-Musawi et al., 2017**) is a solution of the differential equation:

$$\frac{dF_B(t)}{dt} = -\frac{1}{\tau} F_B(t)^n \qquad (1)$$

Where $F_B(t)$ is the cumulative Burr XII distribution function where the variable in this context is the time $(t)$, $n$ is the fractional reaction order parameter; and $\tau$ is a characteristic time which the scale of the process.

Equation (1) solutions are depending on the initial and limits conditions. For a case of decay or a relaxation, this equation becomes:

$$F_B(t) = [1 + (n-1)(\tfrac{t}{\tau})]^{-\frac{1}{n-1}} \qquad (2)$$

Where: $F_B(0) = 1$ and $F_B(\infty) = 0$

Or for an increasing function (sorption until saturation), Equation (1) becomes:

$$F_B(t) = 1 - [1 + (n-1)(\tfrac{t}{\tau})]^{-\frac{1}{n-1}} \qquad (3)$$

where: $F_B(0) = 0$ and $F_B(\infty) = 1$

For both cases, the probability density ($f_B(t)$) and the survival function (or relaxation function, $S_B(t)$) are given by Equations (4 and 5), respectively:

$$f_B(t) = \frac{1}{\tau}[1 + (n-1)\left(\tfrac{t}{\tau}\right)]^{-1-\frac{1}{n-1}} \qquad (4)$$

$$S_B(t) = [1 + (n-1)\left(\tfrac{t}{\tau}\right)]^{\frac{1}{n-1}} \qquad (5)$$

For *n=1*, one recovers the exponential function; for $t \to 0$, one has a linear behavior in $t$, and for $t \to \infty$, one has $S_B(t) \sim t^{-\frac{1}{n-1}}$.

If $f_B(0) = m$ or $f_B(\infty) = m$, eqs. (2-5) have to be multiplied by $m$. Where $m$ is the maximum quantity involved in the kinetics process at $t = 0$ or $t = \infty$.

In order to account for the initial power law behavior for small time observed in most complex systems (**Kopelman, 1988**), an exponent $\alpha$ has been added to the variable of the original Burr function. It is the so-called generalized Burr XII function which due to its extensive use in economy (the Burr-Singh-Maddala function) (**McDonald, 1984; Park and Bera, 2007**) and climatology (**Papalexiou and Koutsoyiannis, 2012; Shao et al., 2004**) is now referred as simply as the Burr XII function. When the variable is the time, this corresponds to the introduction in eqs. (1-5) of a fractal time ($t^a$). Thus, the eqs.(2-5) are now written as:

$$F_{Ba}(t) = 1 - [1 + (n-1)\left(\frac{t}{\tau}\right)^a]^{-\frac{1}{n-1}} \qquad (6)$$

$$f_{Ba}(t) = \frac{a}{\tau}\left(\frac{t}{\tau}\right)^{a-1}[1 + (n-1)\left(\frac{t}{\tau}\right)^a]^{-1-\frac{1}{n-1}} \qquad (7)$$

$$S_{Ba}(t) = [1 + (n-1)\left(\frac{t}{\tau}\right)^a]^{-\frac{1}{n-1}} \qquad (8)$$

$F_{Ba}(t)$ can be formally written as solution of the "fractal" differential equation (**Brouers and Sotolongo-Costa, 2006; Brouers, 2014a**):

$$\frac{dF_{Ba}(t)}{dt^a} = -\frac{1}{\tau}F_{Ba}(t)^n \qquad (9)$$

or

$$\frac{dF_{Ba}(t)}{dt} = -\frac{at^{a-1}}{\tau}F_{Ba}(t)^n \qquad (10)$$

Alternatively, (9) and (10) can be combined as shown in Equation (11) below (**Singh and Maddala, 1976; Brouers, 2015**):

$$\frac{dF_{Ba}}{dx} = f(x) = g(x)F_{Ba}(x)(1 - F_{Ba}(x)) \qquad (11)$$

where the function $g(x) = \frac{\tilde{g}(x)}{x}$ with $\tilde{g}(x) = \dfrac{\left(a\left(\frac{x}{b}\right)^\alpha\right)}{\left(1+(n-1)\left(\frac{x}{b}\right)^\alpha\right)\left(1-\left(1+(n-1)\left(\frac{x}{b}\right)^\alpha\right)^{-\frac{1}{n-1}}\right)}$

The function $\tilde{g}(x)$ varies slowly from $\alpha$ to $\alpha/c$. One has: $\tilde{g}(x) \to \alpha$ when $x \to 0$ and $\tilde{g}(x) \to \frac{a}{c}$, when $x \to \infty$.

The differential equation describes a birth and death process modulated by a quasi-hyperbolic function which expresses the irreversibility of the kinetics. Equations (9-11) can be modified to represent more complex systems as this is the case in the field of epidemiology.

The cumulative distribution function $F_{B\alpha}(x)$ exhibits asymptotically two power laws: one for $x \to 0$, $F_{B\alpha}(x) \to x^a$, and one for $x \to \infty$, $F_{B\alpha}(x) \to x^{-\alpha/(n-1)}$. It has a limited number of finite moment depending on the value of $\alpha/(n-1)$.

The two asymptotic behaviors of the survival function are:

$$S_{B\alpha}(t) \to \left(\frac{t}{\tau}\right)^\alpha \qquad \text{For } t \ll \tau \qquad (12)$$

$$S_{B\alpha}(t) \to \left(\frac{t}{\tau}\right)^{-\left(\frac{\alpha}{n-1}\right)} \qquad \text{For } t \gg \tau \qquad (13)$$

When $\left(\frac{\alpha}{n-1}\right) < 1$, , the Burr XII function belong to the basin of attraction of the heavy tail Lévy distributions. Equation (6) can be written as a deformed Weibull function known as *q*-Weibulll function (*q=n-1*) (**Picoli et al., 2003**):

$$F_{Ba}(t) = 1 - \exp_q\left(-\left(\frac{t}{\tau}\right)^\alpha\right) \qquad (14)$$

Where $\exp_q(x)$ is the $q$-deformed exponential $(1 - qx)^{-\frac{1}{q}}$

The differential eqs. (9 and 10) can be rewritten in the form of a simple first order differential equation:

$$\frac{dF_{B\alpha}(t)}{dt} = -R(t)F_{B\alpha}(t) \qquad (15)$$

With a time dependent rate or hazard function (in reliability theory) or intensity of transition (in relaxation theory), $R(t)$ is given by (**Brouers and Sotolongo-Costa, 2006; Brouers, 2014a; Jurlewicz and Weron, 1999; Stanislavsky and Weron, 2017**).

$$R(t) = -\frac{d}{dt}\ln(S_{B\alpha}(t)) = \frac{1}{\tau(t)} = \frac{1}{\tau}\frac{\left(\frac{t}{\tau}\right)^{\alpha-1}}{\left(1+(n-1)\left(\frac{t}{\tau}\right)\right)} \qquad (16)$$

The proposed kinetics equation based on this theory in the field of sorption has been named $BSf(n, \alpha)$ (eq.(17)) or sometimes $BSW(t)$.

$$BSf(n, \alpha) \equiv q(t) = q_m\left[1 - \left(1 + (n-1)\left(\frac{t}{\tau}\right)^\alpha\right)^{-\frac{1}{n-1}}\right] \qquad (17)$$

Where $q(t)$ and $q_m$ are the time dependent and maximum sorbed quantities, respectively (mg/g); $n$ is the fractional order of the reaction; $\alpha$ is the fractal coefficient expressing macroscopically the complexity of the sorbent-sorbate couple, and $\tau$ is a characteristic time (min). In addition, the half time $\tau_{50\%}$ (min) corresponding to $q(t) = 0.5\, q_m$ is given by:

$$\tau_{50\%} = \tau\left(\frac{(0.5)^{-n+1}-1}{n-1}\right)^{\frac{1}{\alpha}} \qquad (18)$$

As discuss in references (**Brouers and Sotolongo-Costa, 2006; Brouers, 2014a; Al-Musawi et al., 2017**), from the general equation $BSf(n, \alpha)$ equation, one can derive some of the most used empirical sorption kinetics equations. Which are i) $BSf(1, 1)$ is the pseudo-first order kinetics equation, ii) $BSf(1,\alpha)$ is the fractal pseudo-first order or Weibull-Avrami equation, iii) $BSf(2,1)$ is the pseudo-second order kinetics equation, iv) $BSf(2,\alpha)$ is the Hill (or log-logistic) kinetics equation and, v) $BSf(1.5,\alpha)$ is the Brouers-Gaspard kinetics equation. In practice in sorption problems, as discussed by **Al-Musawi et al. (2017)**, the order *n* is not easily determined due to the scare number of data in the saturation region which cannot provide a precise value of the asymptotic exponent $(\alpha/(n-1))$ and the relevant quantity of the fractal index $\alpha$. As a consequence in most case *BSf(1,a)* and *BSf(2,a)* give very close results and to compare a series of data where some are better represented by *BSf(1,a)* and others by *BSf(2,a)* it has been suggested to use *BSf(1.5,a)* for the whole series (**Al-Musawi et al., 2017**). A similar approach has been introduced in the theory of relaxation (**Brouers and Sotolongo-Costa, 2005; Brouers et al., 2004**) and to model isotherms data (**Brouers et al., 2005; Ncibi et al., 2008; Brouers , 2013; Brouers, 2014b; Brouers and Al-Musawi, 2015; Brouers and Marquez-Montesino, 2016**).

It is interesting to note that when $\alpha = 1$ in equation (8) the BurrXII survival function is the Tsallis density function while in eq.(7) the BurrXII density function is the Tsallis "*escort*" probability in the non-extensive theory (**Tsallis, 2002**).

### 2.2.2 The fractional derivative kinetic equation

Recently in parallel with $BSf(n,\alpha)$ formalism, the fractional calculus has been introduced in the field of relaxation and sorption kinetics in order to introduce formally memory effects (**Tomczak et al., 2013; Kaminski et al., 2016; Vinnett et al., 2015; Friesen et al., 2015; Li et al., 2016; Gorenflo and Mainardi, 2007; Garrappa et al.,**

**2016; Khamzin et al., 2013**). The fractional derivative kinetics equation is a formal generalization of the first order differential equation (19) governing the exponential decay with $n = 1$:

$$\frac{dS(t)}{dt} = -\frac{1}{\tau} S(t) \qquad (19)$$

Where $t \geq 0$ and $S(0^+) = 1$.

Whose solution is shown in equation (20) below:

$$S(t) = exp\left(-\frac{t}{\tau}\right) \qquad (20)$$

Fractional derivatives have been proposed in two forms, using the Riemann-Liouville fractional derivative or the second the Caputo fractional derivative (eqs.: 21 and 22) (**Li et al., 2016; Gorenflo and Mainardi, 2007**).

$$\frac{dS(t)}{dt} = -D_t^{1-\alpha} S(t), \qquad t \geq 0 \qquad with \qquad S(0^+) = 1 \qquad (21)$$

or

$$_cD_t^\alpha S(t) = -S(t) \qquad t \geq 0 \qquad with \qquad S(0^+) = 1 \qquad (22)$$

As far as we are concerned, the two forms are equivalent since the Laplace transform of both solutions are equivalent (**Tomczak et al., 2013**). The fractional derivative of order $\alpha > 0$ in the Riemann-Liouville sense is defined as the operator:

$$D_t^\alpha J_t^\alpha = I \qquad \alpha > 0 \qquad (23)$$

Where $J_t^\alpha$ is the Riemann-Liouville fractional integral. For any $\alpha > 0$, this fractional integral is defined as:

$$J_t^\alpha \varphi(t) = \frac{1}{\Gamma(\alpha)} \int_0^\infty (t - t')^{\alpha - 1} \varphi(t') dt \qquad (24)$$

Where $\Gamma(\alpha) = \int_0^\infty e^{-u} u^{\alpha-1} du$, is the Gamma function. For the existence of the integral (24), it is sufficient that the function $\varphi(t)$ be locally integrable in $R^+$ and for $t \to 0$ behaves like $\varphi(t^{-\nu})$ with a number $\nu < \alpha$.

One of the most useful properties of the Riemann-Liouville fractional derivative is given by its Laplace transform as:

$$\mathcal{L}\{D_t^\alpha \varphi(t); s\} = s^\alpha \tilde{\varphi}(s) \qquad (25)$$

The sign $\mathcal{L}$ means the Laplace transform.

If we introduce a relaxation time $\tau^*$ as a time scale, Equation (21) becomes

$$\frac{dS(t)}{dt} = -\tau^{*-\alpha} D_t^{1-\alpha} S(t), \qquad t \geq 0 \qquad \text{with} \qquad S(0^+) = 1 \qquad (26)$$

On the other hand, as it is well known, the Laplace transform of the Mittag-Leffler (*ML*) function is given by (**Li et al., 2016; Gorenflo and Mainardi, 2007; Garrappa et al., 2016**)

$$\mathcal{L}\{E_\alpha(-(t/\tau^*)^\alpha); s\} = \frac{1}{s((s\tau^*)^{-\alpha}+1)} \qquad (27)$$

Which allows us to write the fractional derivative survival function or relaxation function as:

$$S_{FD}(t) = E_\alpha\left(-\left(\frac{t}{\tau^*}\right)^\alpha\right) \qquad t \geq 0 \qquad 0 < \alpha \leq 1 \qquad (28)$$

Where:

$$E_\alpha(z) = \sum_{k=0}^\infty \frac{z^k}{\Gamma(1+\alpha k)} \qquad (29)$$

For $\alpha = 1$, one recovers an exponential (Debye) decay, $f_{FD}(t)$ is given by:

$$f_{FD}(t) = \frac{t^{\alpha-1} E_{\alpha\alpha}[-(t/\tau^*)^\alpha]}{(\tau^*)^\alpha} \qquad (30)$$

Where $E_{\alpha\beta}[-(t/\tau^*)^\alpha]$ is the generalized two parameters *ML* function as shown in Eq.(31):

$$E_{\alpha.\beta}(z) = \sum_{k=0}^{\infty} \frac{z^k}{\Gamma(\beta+\alpha k)} \quad \text{and} \quad E_{\alpha.1}(z) = E_\alpha(z) \tag{31}$$

The rate here is given by:

$$R(t) = \frac{f_{FD}(t)}{S_{FD}(t)} = \frac{t^{\alpha-1} E_{\alpha\alpha}[-(t/\tau^*)^\alpha]}{(\tau^*)^\alpha E_\alpha[-(t/\tau^*)^\alpha]} \tag{32}$$

It is interesting to observe that this solution corresponds to the Cole-Cole theory of relaxation. Indeed, **Stanislavsky and Weron, (2017)**; **Li et al., (2016)**; **Gorenflo and Mainardi, (2007)**; and **Garrappa et al., (2016)** have obtained the same result by Laplace transforming the empirical Cole-Cole susceptibility function used in the frequency domain. In the theory of relaxation, the Cole-Cole expression of the susceptibility in the frequency domain which is used to interpret experimental data, is obtain by inserting an empirical exponent in the classical (exponential) Debye form ($\alpha = 1$):

$$\chi_{CC}(s) = \frac{1}{1+(s\tau^*)^\alpha} \quad \text{with} \quad s = i\omega \tag{33}$$

where $\tau^*$ is a characteristic time corresponding to the peak in the susceptibility.

The Cole-Cole response function $f_{CC}(t)$ is given by the inverse Laplace transform of $\chi_{CC}(s)$:

$$f_{CC}(t) = \mathcal{L}^{-1}\left(\frac{1}{1+(s\tau^*)^\alpha}\right) = \frac{dS_{CC}(t)}{dt} = \frac{1}{\tau^*}\left(\frac{t}{\tau^*}\right)^{\alpha-1} E_{\alpha\alpha}\left(-\left(\frac{t}{\tau^*}\right)^\alpha\right) \tag{34}$$

And the relaxation function:

$$S_{CC}(t) = \mathcal{L}^{-1}\left(\frac{1}{s}\left(1 - \frac{1}{1+(s\tau^*)^\alpha}\right)\right) = \mathcal{L}^{-1}\left(\frac{1}{s}\left(\frac{(s\tau^*)^\alpha}{1+(s\tau^*)^\alpha}\right)\right) = E_\alpha\left(-\left(\frac{t}{\tau^*}\right)^\alpha\right) \tag{35}$$

Therefore, in the framework of the fractional derivative theory, the sorption kinetics equation which has been used in sorption and to be compared with (17) is:

$$q_{FD}(t) = q_m[1 - E_\alpha\left(-\left(\frac{t}{\tau^*}\right)^\alpha\right)] \qquad t \geq 0 \qquad 0 < \alpha \leq 1 \qquad (36)$$

Using the asymptotic behavior of the *ML* function is given by function we have:

$$q_{FD}(t) \rightarrow q_m \frac{1}{\Gamma(\alpha+1)} \left(\frac{t}{\tau^*}\right)^\alpha \qquad t \ll \tau \qquad (37)$$

$$q_{FD}(t) \rightarrow q_m(1 - \frac{1}{\Gamma(\alpha+1)} \left(\frac{t}{\tau^*}\right)^{-\alpha}) \qquad t \gg \tau \qquad (38)$$

The two functions eq. (36) and eq. (17) have the same asymptotic behavior for $t \rightarrow 0$, if $\tau^* = \tau\Gamma(1+\alpha)^{-\frac{1}{\alpha}}$. They have the same asymptotic behavior for $t \rightarrow \infty$ when $n=2$ in $BSf(n,\alpha)$. This is the form used recently in the literature by (**Friesen et al., 2015; Kaminski et al., 2016; Li et al., 2016; Gorenflo and Mainardi, 2007; Garrappa et al., 2016**). Comparing both equations, one can notice that eq.(17) has two complexity parameters $\alpha$ and *n* while Equation (36) has only one. It is worth to mention that some data were presented in their corresponding references as a percentage uptake ($q\%$ that calculated using the following equation:

$$q\% = \left(\frac{q(t)}{q_m}\right) \times 100 \qquad (39)$$

## 3. Results and discussion

In this study, a detailed comparison of the two methods using some recent published data was presented. The full calculations have been done with the nonlinear fitting programming of Mathematica software (version 10) which includes the computation of the ML functions. First we will proceed with a formal comparison determining, for a chosen value of $\alpha$ in the *ML* and the $BSf(n,\alpha)$ formulas, the value of $n$ which yields the same calculated kinetics. It appears that for a given common value of $\alpha$ the fractional solution almost coincides numerically with the fractal solution having $\alpha$ value of $n$ such that $1 \leq n \leq 2$, (see Fig.1 a-c). For the comparison, to have the same asymptotic behavior for $t \rightarrow 0$, we have use for the scaling factors $\tau = 1$ and $\tau^* = \Gamma(1+\alpha)^{-1/\alpha}$. As recalled earlier in sorption problems $n$ is not easily determined, $\alpha$ being the relevant macroscopic parameter, we can therefore anticipate that in practice both methods can give very similar representation of the experimental curve. For a common value of $\alpha$ in $ML(\alpha)$ and $BSf(n, \alpha)$ determination of the value of $n$ in $BSf(n, \alpha)$ yielded the same computed kinetics. We have therefore, in this particular example, $ML(0.25) \cong BSf(2, 0.25)$, $ML(0.50) \cong BSf(1.75, 0.50)$, $ML(0.75) \cong BSf(1.75, 0.75)$.

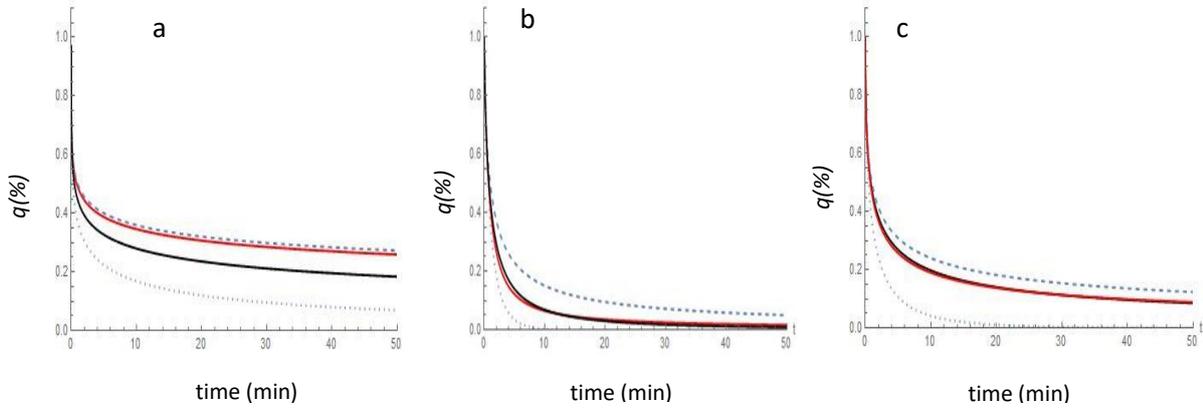

Fig. 1. Calculations of $\alpha$ in the $ML$ and $BSf(n, \alpha)$ formulas; (a) [red solid: ML(0.25); dotted: $BSf(1, 0.25)$ ; dashed: $BSf(2, 0.25)$ ; black solid: $BSf(1.75, 0.25)$]; (b) [red solid: ML(0.5); dotted: $BSf(1, 0.50)$; dashed: $BSf(2, 0.50)$ ; black solid: $BSf(1.75, 0.50)$], and (c) [red solid: ML(0.75); dotted: $BSf(1, 0.75)$ ; dashed: $BSf(2, 0.75)$ ; black solid: $BSf(1.75, 0.75)$]

We are now going to compare the results of both methods using seven set of data from the recent literature and some unpublished data, the results are tabulated in Tables (1- 7). They will confirm the conclusion of the theoretical comparison. One can compare the results of *sets 1, 2, 3 and 4*, there is practically no difference between the results for the three characteristic coefficient and using $BSf(n,\alpha)$ (with $n=1$) and eq.36. The introduction of an alternating current does not change the sorption power but increases the velocity of the reaction. This is due to an increase of the fractal coefficient and the corresponding decrease of the half-time by a factor of order 3. The calculations with both methods quantize what is apparent in the data.

We have treated the data of reference (**Vinnett et al., 2015**) calculating the percentage of copper uptake (*set* 5). Here the two methods described in section 2.2.1 and 2.2.2, provide similar results. In that paper a different fractional equation has been used introducing a three parameter function different from the ML functions. The comparison shows that their parameter $\alpha$ does not have a clear fractal or asymptotic behavior interpretation.

**Table 1:** Results of the modeling kinetics of *set* 1 data

| Kinetic model | $q\%$ | $\alpha$ | $\tau$ (min) | $\tau_{0.5}$ (min) | $R^2$ |
|---|---|---|---|---|---|
| *Initial TC conc.= 10 mg/L* | | | | | |
| Weibull ($BSf(1,a)$) | 75.3 | 0.80 | 41.0 | 25.9 | 0.9973 |
| Mittag-Leffler ($\alpha$) | 71.6 | 0.79 | 37.6 | | 0.9975 |
| *Initial TC conc.= 20 mg/L* | | | | | |
| Weibull ($BSf(1,\alpha)$) | 89.9 | 0.90 | 39.4 | 26.3 | 0.9974 |
| Mittag-Leffler ($\alpha$) | 71.6 | 0.79 | 39.0 | | 0.9977 |
| *Initial TC conc.= 30 mg/L* | | | | | |
| Weibull $BSf(1,\alpha)$ | 93.9 | 0.88 | 30.8 | 20.3 | 0.9971 |
| Mittag-Leffler ($\alpha$) | 92.2 | 0.86 | 29.3 | | 0.9973 |
| *Initial TC conc.= 40 mg/L* | | | | | |

| Weibull ($BSf(1,\alpha)$) | 97.3 | 0.85 | 28.6 | 18.6 | 0.9971 |
| Mittag-Leffler ($\alpha$) | 98.52 | 0.83 | 27.1 | | 0.9973 |

Table 2: Results of the modeling kinetics of *set 2* data

| Kinetic model | $q_m$ (mg/g) | $\alpha$ | $\tau$ (min) | $\tau_{0.5}$ (min) | $R^2$ |
|---|---|---|---|---|---|
| *Initial indigo carmine conc.= 80 mg/L* | | | | | |
| Weibull ($BSf(1,\alpha)$) | 30.7 | 0.64 | 716.3 | 405.6 | 0.9992 |
| Mittag-Leffler ($\alpha$) | 28.9 | 0.64 | 615.8 | | 0.9991 |
| *Initial indigo carmine conc.= 100 mg/L* | | | | | |
| Weibull ($BSf(1,\alpha)$) | 34.4 | 0.59 | 1223 | 656.4 | 0.9996 |
| Mittag-Leffler ($\alpha$) | 31.3 | 0.58 | 1013 | | 0.9996 |

Table 3: Results of the modeling kinetics of *set 3* data

| Kinetic model | $q_m$ (mg/g) | $\alpha$ | $\tau$ (min) | $\tau_{0.5}$ (min) | $R^2$ |
|---|---|---|---|---|---|
| *Initial indigo carmine conc.= 80 mg/L* | | | | | |
| Weibull ($BSf(1,\alpha)$) | 27.9 | 0.99 | 249.3 | 172.0 | 0.9944 |
| Mittag-Leffler ($\alpha$) | 26.8 | 0.97 | 236.5 | | 0.9950 |
| *Initial indigo carmine conc.= 100 mg/L* | | | | | |
| Weibull ($BSf(1,\alpha)$) | 32.8 | 0.99 | 237.4 | 164.5 | 0.9974 |
| Mittag-Leffler ($\alpha$) | 31.8 | 0.97 | 230.4 | | 0.9980 |

Table 4: Results of the modeling kinetics of *set 4* data

| Kinetic model | $q_m$ (mg/g) | $\alpha$ | $\tau$ (min) | $\tau_{0.5}$ (min) | $R^2$ |
|---|---|---|---|---|---|
| *Initial cyanide conc.= 25 mg/L* | | | | | |
| Weibull ($BSf(1,\alpha)$) | 99.91 | 0.45 | 13.3 | 5.85 | 0.9999 |
| Mittag-Leffler ($\alpha$) | 94.97 | 0.44 | 11.3 | | 0.9999 |
| *Initial cyanide conc.= 50 mg/L* | | | | | |
| Weibull ($BSf(1,\alpha)$) | 73.89 | 0.58 | 9.47 | 5.82 | 0.9998 |
| Mittag-Leffler ($\alpha$) | 71.14 | 0.55 | 8.66 | | 0.9999 |
| *Initial cyanide conc.= 75 mg/L* | | | | | |
| Weibull ($BSf(1,\alpha)$) | 77.03 | 0.32 | 4.63 | 1.47 | 0.9998 |
| Mittag-Leffler ($\alpha$) | 73.43 | 0.31 | 3.75 | | 0.9999 |

**Table 5:** Results of the modeling kinetics of *set 5* data

| **Kinetic model** | $q_m$ (%) | $\alpha$ | $\tau$ (day) | $\tau_{0.5}$ (day) | $R^2$ |
|---|---|---|---|---|---|
| *Initial copper conc.= 10 mg/L* | | | | | |
| Weibull ($BSf(1,\alpha)$) | 95 | 0.75±0.06 | 0.17 | 0.10 | 0.9999 |
| Mittag-Leffler ($\alpha$) | 97 | 0.85±0.06 | 0.17 | | 0.9999 |
| Experimental | | 0.21±0.03 | | | 0.99 |
| *Initial copper conc.= 15 mg/L* | | | | | |
| Weibull ($BSf(1,\alpha)$) | 97 | 0.78±0.08 | 0.21 | 0.13 | 0.9993 |
| Mittag-Leffler ($\alpha$) | 98 | 0.87±0.06 | 0.21 | | 0.9993 |
| Experimental | | 0.21±0.03 | | | 0.99 |

Recently a paper has been published on the sorption of water in cement (Saeidpour, M., Wadsö, L., 2015). The data contain more than hundred kinetics sorption points measured with great precision. In that case, it is possible to determine the exact value of *n* and therefore a comparison between the full *BSf(n,a)* equation and the *ML* form is more accurate (see table 6 and Fif.2) From this table, one can get same conclusions. The two methods can give very close results for the fractal coefficient and the time scale $\tau$. The knowledge of the apparent order *n* yield a better knowledge of the rate as a function of time. Due to the quality of experimental data, we have reported the statistical error (*MSE*) range of the fitting program of Mathematica.

Taking advantage of the great number of experimental points presented by (**Saeidpour and Wadsö, 2015**), we have introduced a further generalization of the fractal kinetics equation by assuming a time dependence of the fractal exponent (eq.40) to take account of the modification of the sorbing conditions as the process evolves in time. In the real space domain, fractal variation with the scale has been introduced in the theory of limited aggregation clusters. For instance, one has observed a change of fractal dimension with time and therefore scale during clustering processes in aggregation phenomena. As usual in the theory of complex systems we have assumed a power law variation of the fractal exponent between its initial value $\alpha_o$ (at $t = 0$) and $\alpha_s$ (at $t=$ saturation time, $t_s$) where the sorption can be assumed to be saturated. A new parameter $\nu$ has been introduced which is a measure of the time evolution of the fractal coefficient $\alpha$. In that case, the best fit is obtained for a variation of $\alpha$ from 0.65 to 0.47.

$$\alpha(t) = \alpha_o + (\alpha_s - \alpha_o)\left(\frac{t}{t_s}\right)^\nu \qquad (40)$$

We have compared with the calculations of Li et al 2016 using *BSf(n, α)* under another confusing name. Because of the great number of data and therefore the highest precision, we have added in the table the error range (Error) in which the parameter around their more probable value could lie using results given by the Mathematica program.

In that case we have also calculated the rate (eq.16) (**Fig. 3**). The *BSf(1.180, 0.655)* rate is closer to Weibull rate. The ML rate is also between *BSf(1, α)* and *BSf(2, α)* rates and has calculated asymptotic behaviors closer to the Hill rate. The reason is that the asymptotic exponents of the survival function $S_{Ba}(t)$ are the same in both cases (see eqs.12-13). The Tsallis or q-exponential kinetics rate where the fractal exponent is 1, is much lower because it neglects an important fractal factor of the kinetics process.

**Table 6:** Results of the modeling kinetics of *set 6* data

| Kinetic model | $q_{max}$ | $\alpha$ | $n$ | $\tau$ (min) | $\tau_{50}$ (min) | $R^2$ | $MSE$ |
|---|---|---|---|---|---|---|---|
| Tsallis: $BSf(n,1)$ | 1.49 | 1 | 6.6 | 10460 | 7250 | 0.999639 | 9x10$^{-5}$ |
| Error | 0.1 | | 0.6 | 505 | | | |
| Weibull ($BSf(1,\alpha)$) | 0.767 | 0.629 | 1 | 15585 | 8699 | 0.999976 | 6x10$^{-6}$ |
| Error | 0.001 | 0.002 | | 64 | | | |
| Hill: $BSf(2, \alpha)$ | 0.918 | 0.736 | 2 | 13190 | 13190 | 0.999933 | 1.7x10$^{-5}$ |
| Error | 0.004 | 0.004 | | 184 | 184 | | |
| $BSf(n, \alpha)$ | 0.793 | 0.655 | 1.18 | 14767 | 9308 | 0.999984 | 4x10$^{-6}$ |
| Error | 0.010 | 0.005 | 0.06 | 114 | | | |
| $BSf(n, \alpha(t))$ where $\nu$=0.5 | 0.821 | 0.65-0.47 | 1.04 | 18237 | | 0.999985 | 4x10$^{-6}$ |
| Error | 0.010 | 0.05 | 0.1 | 7175 | | | |
| Mittag-Leffler ($\alpha$) | 0.862 | 0.672 | | 18288 | | 0.999980 | 5x10$^{-6}$ |
| Error | 0.002 | 0.002 | | 124 | | | |
| **Li et al. 2016** | 0.804 | 0.660 | 1.24 | | | 0.99 | |

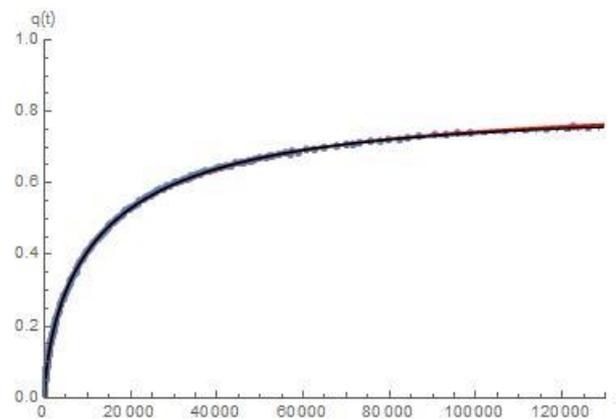

**Fig. 2.** Fit of *set 6* data [dotted dashed: *BSf(1.18,0.655)* and Red solid ML(0.672) ]

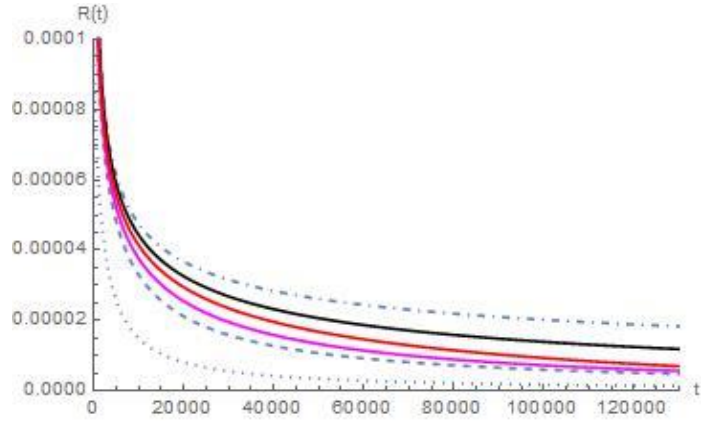

**Fig.3.** Rate of *set 6* data, [dotted dashed: Weibull; black solid *BSf(1.180, 0.655);* red solid: *BSf(n, α(t))*, magenta solid: *ML(0.672)*, dashed: Hill, dotted: Tsallis]

The last set of data is the kinetics of the removal of fluorine from aqueous solutions by clay (**Brouers and Guiza, 2017**). Where the fits give similar result for the *BSf(1, α)* (Weibull-Avrami equation ) and the ML kinetics formula (Table 7).

**Table 7:** Results of the modeling kinetics of *set 7* data

| Kinetic model | $q_m$ (mg/g) | $\alpha$ | $\tau$ (min) | $\tau_{50\%}$ (min) | $R^2$ |
|---|---|---|---|---|---|
| $BSf(1,1)$ | 24.0 | 1 | 5.8 | 4.06 | 0.9977 |
| $BSf(1,\alpha)$ | 26.3 ± 0.7 | 0.47±0.05 | 5.8±0.05 | 2.67 | 0.9998 |
| $BSf(2,1)$ | 26.2 | 1 | 3.58 | 3.58 | 0.9996 |
| $BSf(2,\alpha)$ | 28,8 | 0.67 | 3.65 | 3.65 | 0.9997 |
| ML | 26.3±0.07 | 047 ±0.05 | 5.8 ±0.06 |  | 0.9998 |

## 4. Conclusions

The two methods can provide fits with almost the same precision. Both have a degree of empiricism. *Bf(n,α)* introduces an empirical fractal time and the fractional equation is the Laplace transform of the empirical Cole-Cole expression of the relaxation susceptibility in the frequency domain. We believe the *BSf(n,a)* is in practice more useful in that field when data are small compared with those used in econometrics or life-time studies. The reason of our preference is that, being a well-defined distribution function (BurrXII as well as its approximations Weibull and Log-logistic distributions) and knowing their statistical properties, it is easier to characterize physically and mathematically the sorption properties of a sorbent-sorbate couple. In particular, it yields simple analytical expressions for the rate equation and the half- life time. Another reason, more fundamental from a theoretical point of view is that the $BSf(n, \alpha)$ can be derived from a stochastic description giving a physical meaning to the two coefficient $\alpha$ and $n$ which are related to the multiscale fractal organization and cluster architecture of internal system at the micro- and mesoscopic scales. Another practical advantage of $BSf(n, \alpha)$ is that it is has a simple analytic form contrary to the Mittag-Leffler formula and as a consequence, it can be applied simply to other problems. For that reason the computational time is also at least five times more rapid.